\documentstyle{mn}

\newif\ifAMStwofonts

\def\pcm3{{\rm\thinspace cm^{-3}}}

\def\contcaption{\@conttrue\SFB@caption\@captype}

\def\n_h{{\rm n_{H}}}

\def\NH1{{$N_{\rm HI}~$}}

\def\ga{{\rm\thinspace gauss}}

\def\approxlt{\mathrel{\hbox{\rlap{\lower .5ex \hbox {$\sim$}}
        \raise .15 ex \hbox{$<$}}}}
\def\approxgt{\mathrel{\hbox{\rlap{\lower .5ex \hbox {$\sim$}}
        \raise .15 ex \hbox{$>$}}}}

\def\la{\mathrel{\hbox{\rlap{\hbox{\lower4pt\hbox{$\sim$}}}\hbox{$<$}}}}
\def\ga{\mathrel{\hbox{\rlap{\hbox{\lower4pt\hbox{$\sim$}}}\hbox{$>$}}}}
\newbox\grsign \setbox\grsign=\hbox{$>$} \newdimen\grdimen
\grdimen=\ht\grsign
\newbox\simlessbox \newbox\simgreatbox \newbox\simpropbox
\setbox\simgreatbox=\hbox{\raise.5ex\hbox{$>$}\llap
     {\lower.5ex\hbox{$\sim$}}}\ht1=\grdimen\dp1=0pt
\setbox\simlessbox=\hbox{\raise.5ex\hbox{$<$}\llap
     {\lower.5ex\hbox{$\sim$}}}\ht2=\grdimen\dp2=0pt
\setbox\simpropbox=\hbox{\raise.5ex\hbox{$\propto$}\llap
     {\lower.5ex\hbox{$\sim$}}}\ht2=\grdimen\dp2=0pt
\def\simgreat{\mathrel{\copy\simgreatbox}}
\def\simless{\mathrel{\copy\simlessbox}}

\ifoldfss
  \ifCUPmtlplainloaded \else
    \NewTextAlphabet{textbfit} {cmbxti10} {}
    \NewTextAlphabet{textbfss} {cmssbx10} {}
    \NewMathAlphabet{mathbfit} {cmbxti10} {} 
    \NewMathAlphabet{mathbfss} {cmssbx10} {} 
  \fi
  \ifAMStwofonts
    \ifCUPmtlplainloaded \else
      \NewSymbolFont{upmath} {eurm10}
      \NewSymbolFont{AMSa} {msam10}
      \NewMathSymbol{\upi}     {0}{upmath}{19}
      \NewMathSymbol{\umu}     {0}{upmath}{16}
      \NewMathSymbol{\upartial}{0}{upmath}{40}
      \NewMathSymbol{\leqslant}{3}{AMSa}{36}
      \NewMathSymbol{\geqslant}{3}{AMSa}{3E}

      \let\leq=\leqslant 
      \let\geq=\geqslant 
    \fi
  \fi
\fi 

\ifnfssone
  \newmathalphabet{\mathit}
  \addtoversion{normal}{\mathit}{cmr}{m}{it}
  \addtoversion{bold}{\mathit}{cmr}{bx}{it}
  \newmathalphabet{\mathbfit} 
  \addtoversion{normal}{\mathbfit}{cmr}{bx}{it}
  \addtoversion{bold}{\mathbfit}{cmr}{bx}{it}
  \newmathalphabet{\mathbfss} 
  \addtoversion{normal}{\mathbfss}{cmss}{bx}{n}
  \addtoversion{bold}{\mathbfss}{cmss}{bx}{n}
  \ifAMStwofonts
    \ifCUPmtlplainloaded \else
      \UseAMStwoboldmath
      \makeatletter
      \new@mathgroup\upmath@group
      \define@mathgroup\mv@normal\upmath@group{eur}{m}{n}
      \define@mathgroup\mv@bold\upmath@group{eur}{b}{n}
      \edef\UPM{\hexnumber\upmath@group}
      \new@mathgroup\amsa@group
      \define@mathgroup\mv@normal\amsa@group{msa}{m}{n}
      \define@mathgroup\mv@bold\amsa@group{msa}{m}{n}
      \edef\AMSa{\hexnumber\amsa@group}
      \makeatother
      \mathchardef\upi="0\UPM19
      \mathchardef\umu="0\UPM16
      \mathchardef\upartial="0\UPM40
      \mathchardef\leqslant="3\AMSa36
      \mathchardef\geqslant="3\AMSa3E

      \let\leq=\leqslant 
      \let\geq=\geqslant 
    \fi
  \fi
\fi 

\ifnfsstwo
  \DeclareMathAlphabet{\mathbfit}{OT1}{cmr}{bx}{it}
  \SetMathAlphabet\mathbfit{bold}{OT1}{cmr}{bx}{it}
  \DeclareMathAlphabet{\mathbfss}{OT1}{cmss}{bx}{n}
  \SetMathAlphabet\mathbfss{bold}{OT1}{cmss}{bx}{n}
  \ifAMStwofonts
    \ifCUPmtlplainloaded \else
      \DeclareSymbolFont{UPM}{U}{eur}{m}{n}
      \SetSymbolFont{UPM}{bold}{U}{eur}{b}{n}
      \DeclareSymbolFont{AMSa}{U}{msa}{m}{n}
      \DeclareMathSymbol{\upi}{0}{UPM}{"19}
      \DeclareMathSymbol{\umu}{0}{UPM}{"16}
      \DeclareMathSymbol{\upartial}{0}{UPM}{"40}
      \DeclareMathSymbol{\leqslant}{3}{AMSa}{"36}
      \DeclareMathSymbol{\geqslant}{3}{AMSa}{"3E}

      \let\leq=\leqslant 
      \let\geq=\geqslant 
    \fi
  \fi
\fi 

\ifCUPmtlplainloaded \else
  \ifAMStwofonts \else 
    \def\upi{\pi}
    \def\umu{\mu}
    \def\upartial{\partial}
  \fi
\fi

\title{The missing M dwarfs.}
\author[P. D. Dobbie et al.]
       {P.\,D. Dobbie$^{1}$,   D.\,J. Pinfield$^{2}$, R.\,F. Jameson$^{1}$, S.\,T. Hodgkin$^{3}$ \\
       1. XROA Group, Dept. of Physics \& Astronomy, University of Leicester, University Road, Leicester LE1 7RH \\
2. Astrophysics Research Institute, Liverpool John Moores University, Egerton Wharf, Birkenhead, CH41 1LD \\
3. Institute of Astronomy, Madingley Road, Cambridge CB3 0HA}
 
\date{Accepted 1988 December 15.
      Received 1988 December 14;
      in original form 1988 October 11}

\pagerange{\pageref{firstpage}--\pageref{lastpage}}
\pubyear{1994}

\begin{document}

\maketitle

\label{firstpage}

\begin{abstract}
We present evidence which indicates the luminosity functions of star forming regions, open star clusters and the field dips between spectral types M7-M8. We attribute this to a sharp local drop in the luminosity-mass relation and speculate that this is caused by the beginning of dust formation in the atmospheres of objects in this effective temperature regime. This effect is not predicted by the current generation of low mass stellar/substellar evolutionary models. If our interpretation is correct then this result has important implications for investigations concerned with the mass functions of star forming regions and young open star clusters. For example, it suggests that some brown dwarfs have higher masses than previously thought.      

\end{abstract}

\begin{keywords}
stars: late-type, low-mass, brown dwarfs, atmospheres -- infrared: stars.
\end{keywords}

\section{Introduction}

Over the last few years deep far-red optical and infrared surveys of young open clusters and star forming regions have revealed large populations of brown dwarfs (e.g Bouvier et al. 1998, Pinfield et al. 2000, Magazzu et al. 1998, Luhman et al. 1999, 2000, Lodieu et al. 2002). Indeed, several of the most recent investigations may have unearthed significant numbers of free-floating objects with masses less than the deuterium burning limit (M$\leq0.012$M$_{\odot}$; e.g $\sigma-$Orionis, Zapatero-Osorio et al. 2000, $\theta-$Orionis, Lucas \& Roche 2000, Muench et al. 2002), some possibly having masses as low as M$\sim0.006$M$_{\odot}$ (Lucas \& Roche 2001). The form of the initial mass function (IMF; dN/dm) in star forming regions, young open clusters and in the field, across and below the stellar/substellar boundary and down to at least 0.03M$_{\odot}$, is found to be well matched by a slowly rising powerlaw model with an index in the range $\alpha\sim0-1.5$ (e.g Bouvier et al. 1998, Luhman et al. 1999, Hambly et al. 1999, Reid et al. 1999, Bejar et al. 2001, Dobbie et al. 2002). 

However, there is now some evidence to suggest that below
M$\sim0.03$M$_{\odot}$ the IMF departs from this simple
description. Two independent studies of the $\theta-$Orionis star
forming region appear to show a significant secondary peak in the IMF
close to the deuterium burning limit (Muench et al. 2002, Lucus \&
Roche 2000). The IMF of $\sigma-$Orionis may also rise steeply at
these masses (Bejar et al. 2001). We note that standard star formation
theories (e.g Boss 1998, Low \& Lynden-Bell 1976) predict that the
minimum mass of gas which is unstable against collapse to stellar
densities is of the order 0.01M$_{\odot}$ (e.g Silk 1977). Muench et
al. (2002) have suggested that another formation mechanism may
contribute to the IMF at these very low masses to produce the 
observed secondary peak.  
   
Of course, derivations of the IMF are dependent on the robustness of the conversion from the observational to the theoretical plane. Studies focusing on the very youngest star clusters and star forming regions are undoubtedly complicated by residual dust and gas along the lines of sight and the presence of T-Tau disks (e.g Lucas \& Roche 2000). There are also uncertainties in the theoretical models, particularly at these youngest ages. For example, at ages of less than a few Myrs, predicted T$_{\rm eff}$s and luminosities are found to be highly sensitive to the choice of initial conditions and the adopted treatment of convection (e.g Baraffe et al. 2002). Furthermore, there is now little doubt that the spectral energy distributions of the coolest stellar and substellar objects are strongly influenced by the presence of dust grains in their atmospheres (e.g Tsuji et al. 1996, Jones \& Tsuji 1997, Chabrier et al. 2000, Marley et al. 2002). While the observed trends in colours of the late-M dwarfs through the L dwarfs to the T dwarfs can now be qualitatively reproduced by the latest model atmospheres which incorporate a treatment for both the condensation and sedimentation of species such as Al$_{2}$O$_{3}$, MgSiO$_{3}$ and Fe (e.g Tsuji et al. 2001), the theory of dust grain formation and gravitational settling in stellar photospheres is still very much in its infancy. The effects of these processes on the evolution of substellar objects have not yet been investigated in a fully self-consistent manner.

\begin{figure}
\vspace{17pc}
\includegraphics{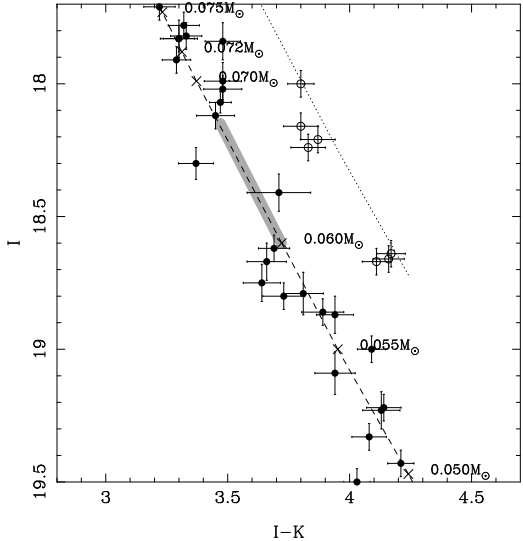}
\caption{A subsection of $I,I-K$ plot for the Pleiades presented in Jameson et al. (2002). Candidate low mass members with photometry consistent with being single and binary objects are shown by the filled and open circles respectively. The 125Myr NEXTGEN isochrone with discrete mass points marked (crosses) and a corresponding sequence for equal mass binaries are also shown (dashed and dotted lines respectively). The sparsely populated region of the cluster sequence between $3.5\leq I-K \leq3.7$ is highlighted (thick grey line).}
\label{plgap}
\end{figure}

In the next section of this work we present evidence for a dip in the
luminosity function (LF) between spectral types M7-M8. In Section 3 we
attribute this to a sharp local change in the form of the luminosity-mass relation. We construct semi-empirical magnitude-mass relations for the Pleiades and $\sigma-$Orionis clusters using the available observational data. In Section~4 we speculate that the change is due to the onset of dust formation in the atmospheres of cool stellar and substellar objects and discuss the implications of our result for current investigations concerned with determining the form of the IMF of star forming regions and young open clusters. It should be noted that throughout this work, we refer to $I_{\rm C}$ as $I$.

\section{Evidence for the missing M dwarfs}

\subsection{The Pleiades}

In our recent work concerned with placing stringent constraints on the
core radius of the substellar members of the Pleiades and on the shape
of the cluster mass function across and below the stellar/substellar
boundary we have compiled a large, complete, magnitude limited sample
of candidate brown dwarfs (Jameson et al. 2002). As part of this
effort we have constructed an $I,I-K$ colour-magnitude diagram (CMD)
using 48 candidate substellar Pleiads which has revealed an apparent
 sparsity of cluster members with colours $3.5\leq I-K \leq3.7$ 
(or $2.5\simless I-J \simless2.7$
from a comparison with the colours of field dwarfs). Using data from
Pinfield et al. (2002, in prep) and Tables~2 and 3 of Jameson et
al. (2002) we
 replot the relevant section of the
CMD in Figure~1 where candidate single and unresolved binary low mass
members are shown as filled and open circles respectively. The
location of the 125Myr NEXTGEN isochrone (Baraffe et al. 1998) for solar metalicity and the
corresponding equal mass binary track, after correcting for
extinction, reddening and a distance of A$_{I}=0.07$, E($I-K$)$=0.06$
and (m-M)$_{0}=5.53$ (Crawford \& Perry 1976) respectively are also
shown (dashed and dotted lines respectively; see arguments in Dobbie
et al. 2002 for our choice of cluster distance).  It can be seen that the 
region of the cluster sequence which we find to be sparsely populated 
(highlighted by the gray shading) spans $\sim0.5$ magnitudes at
$I$. Available infrared data (Jameson et al. 2002; Pinfield et
al. 2002, in prep.) reveal it
 to extend over $\sim0.3$ magnitudes at $J$, $H$ and $K$ 
respectively, suggesting it spans $\sim0.3$ magnitudes in
bolometric luminosity. 

The $I-$band LF over the interval $17.7\leq I \leq 19.50$ for the 7.6
sq. degrees of the Pleiades included in the Jameson et al. (2002)
study is shown in Figure~2. Note that those candidates lying close to
the binary sequence have been excluded from this determination (see
later). Excluding the magnitude bin which has an obvious  deficit of
members, we have modelled log N versus $I$ with a powerlaw (see Muench
et al. 2002 for justification), represented in Figure~2 by the dashed
line. A $\chi^{2}$ test indicates that the observed LF, including the
bin with the deficit of members, is inconsistent with this model at a
level of confidence $\approx90$\% (for the adopted bin width of 0.45
magnitudes). We note that the colour, $I-K\approx3.6$, at the midpoint
of the LF dip corresponds to a spectral type of M7-M8 (Leggett 1992) and an T$_{\rm eff}\approx2700$K (NEXTGEN model). 

\begin{figure}
\vspace{17pc}
\includegraphics{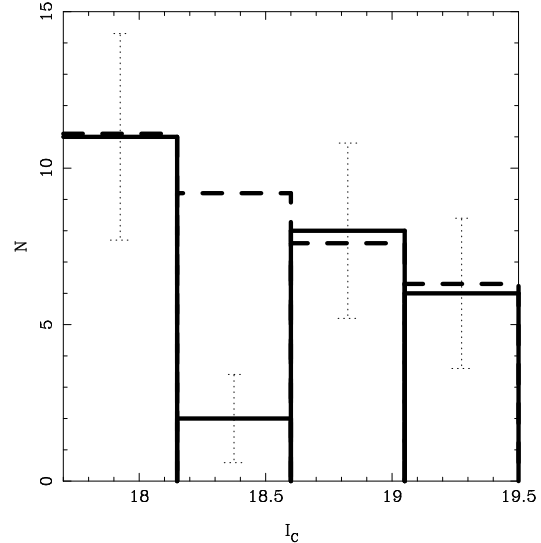}
\caption{The $I-$band LF, in the interval $17.7\leq I \leq 19.5$, of the candidate Pleiads (excluding likely binaries) from the 7.6 sq. degree region of the cluster included in the Jameson et al. (2002) study. A powerlaw model of the LF is also shown (heavy dashed line; see text for details). We find the discrepancy between the observed and model LF to be significant at a level of confidence $\approx90$\%.}
\label{pllum}
\end{figure}

\subsection{$\theta-$ and $\sigma-$Orionis}

Both the $\sigma-$Orionis star cluster and the $\theta-$Orionis
(Trapezium) star forming region, which lie at $\sim400$pc (e.g Blaauw
1991, Warren \& Hesser 1978), are considerably younger than the
Pleiades with estimated ages of 1.7-7 Myrs and 0.3-2 Myrs respectively
(e.g Ali \& Depoy 1995, Hillenbrand 1997, Brown, de Geus \& de Zeeuw
1994, Warren \& Hesser 1978). Bejar et al. (2001) present the results
of a 847 square arcmin $IZJ$ band survey of the $\sigma-$Orionis
region, which is complete to $I=21.5$ and has unearthed 64 candidate
members. Their Figure~1, the $I,I-J$ CMD for the candidates, shows a
sparsity of objects with $2.6\leq I-J \leq3.0$, virtually coincident
in colour with the LF dip observed in the sequence of the
Pleiades. Lucas \& Roche (2000) present the results of an $I$, $J$ and
$H$ band survey of $\theta-$Orionis complete to $H\approx18.0$, in
which 515 sources were detected in a 33 sq. arcminute region. The vast
majority of these are believed to be members of the Trapezium. A
$J-H$, $I-J$ colour-colour plot of these sources, constructed using
dereddened photometry, also reveals a sparsity of objects in the range
$2.6\leq I-J \leq 2.8$ (see their Figure~3). Their $I-$band
LF for the
Trapezium region is falling into the substellar regime but appears to
rise again towards a secondary peak at M$_{I}\approx10.5$
($H\approx16.5$, $J\approx17.5$). The $K-$band LF derived from an
independent near-IR survey of the region similarly displays such a fall in the substellar regime, before apparently rising towards a secondary peak at $K\approx15.5$. Muench et al. (2002), who argue that the theoretical luminosity-mass relation is reasonably robust in the substellar regime, interprete this apparent secondary peak as a rise in the IMF near the deuterium burning limit. 

\begin{figure}
\vspace{17pc}
\includegraphics{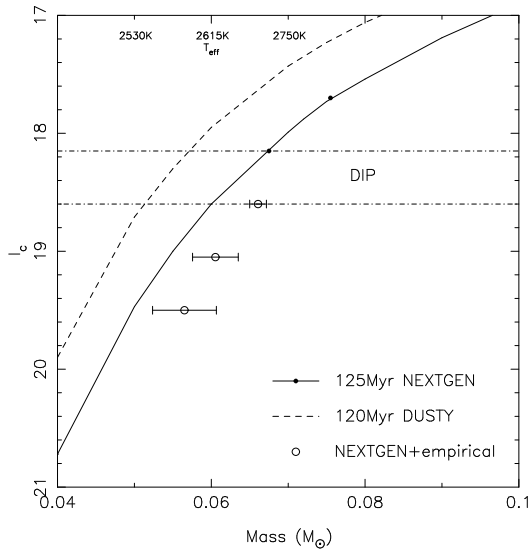}
\caption{The magnitude-mass relation for the Pleiades
low-mass stars and brown dwarfs as determined from the observations. The purely
theoretical 125Myr NEXTGEN and 120 Myr DUSTY relations of Baraffe et
al. (1998) and Chabrier et al. (2000) are also shown (solid and dashed
lines respectively). The location of the dip in the LF is highlighted
(dotted lines).}
\label{pllm}
\end{figure}

\begin{figure}
\vspace{17pc}
\includegraphics{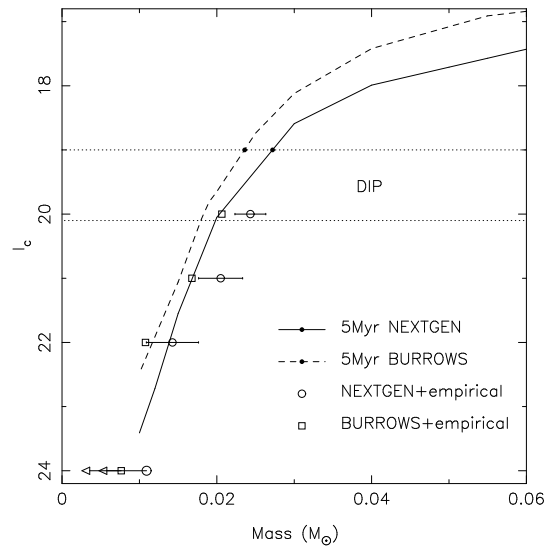}
\caption{The magnitude-mass relation for the brown dwarfs of
$\sigma-$Orionis as determined from the observations. The purely
theoretical 5Myr relations of Baraffe et al. (1998) and Burrows et
al. (1997) are also shown (NEXTGEN - solid line, Burrows - dashed
line). Note that for each calculation, the bottom point is an upper
limit. The location of the LF dip is again
highlighted (dotted lines).}
\label{slom}
\end{figure} 
 
\subsection{Other clusters and the field}

Further evidence for drops in LFs in this T$_{\rm eff}$ regime can
be found elsewhere. For example,  Barrado y Navascu\'es et al. (2001)
present the results of a VRIZ survey, complete to $I\approx18$, of the
53Myr old open cluster IC2391. In a 2.5 square degree area, they
identify 132 probable and possible cluster members, although as
follow-up data are available for only a limited number of these
$\sim25$\% are likely to be non-members. From their Figure~3, the $I,
R-I$ CMD, it can be seen that the number of candidates drops
dramatically just above the completeness limit of their survey, before
increasing once again at $I\simgreat19.5$. The 50Myr NEXTGEN model of
Barraffe et al. (1998) indicates that at the cluster distance of 155
pc (Barrado y Navascu\'es et al. 2001) a member of magnitude
$I\approx18$ has T$_{\rm eff}\approx2750$K. This is consistent with
the theoretical T$_{\rm eff}$ of the top edge of the sparsely
populated region in the Pleiades sequence. The $I-$band LF of the 90Myr old
$\alpha$-Per cluster similarly shows a dip centered around
M$_{I_{c}}\approx12.5$ (Barrado y Navascu\'es, private comm.; Stauffer et
al. 1999), while
 Figure 9 of Luhman (1999), a luminosity-T$_{\rm eff}$ plot, hints at
a sparsity of 
objects between spectral types M7-M8 in IC348. Reid \& Cruz (2002) have derived
colour-absolute magnitude diagrams for a sample of field dwarfs lying
within 20pc of the Sun. Examination of their Figure~11, the
M$_{I}$,$I-J$ plot, also reveals a sparsity of objects in the range
$2.6\leq I-J \leq3.0$. It is clear that with the small number
of late-M type stellar/substellar objects in each of these studies,
and in the IC2391 work the potential level of contamination and the
location of completeness limit of the data with respect to our claimed
dip, individually these results are of low statistical significance. 
Nevertheless, when all the available evidence is taken together we 
believe the case for a drop in the bolometric LF between M7-M8 is compelling.  

\section{Masses across and below the gap}

We have argued above for the existence of a dip in the bolometric 
LF of star forming regions, young clusters and the field near T$_{\rm
eff}\approx2700$K. These populations span a wide range of ages, hence
the dip covers a large range of masses, arguing against it being
related to fine structure in the IMFs. Instead, the obvious way to
explain this is a sharp fall in the luminosity-mass relation between
spectral types M7-M8. The LF can be written as in Equation 1, where
$dN/dm$ is the MF and $dM/dm$ is the slope of the luminosity-mass
relation where M is the bolometric magnitude and m the mass.

\vspace{0.1cm}
\begin{center}
\begin{equation}
dN/dM = \frac{dN}{dm} / \frac{dM}{dm}
\end{equation}
\end{center} 
\vspace{0.1cm}

Logically, if $dM/dm$ increases, the LF will drop. Indeed, we believe
 that what has recently been interpreted as structure in the Trapezium
 IMF (Meunch  et al. 2002) in fact arises from the feature in the
 luminosity-mass relation reported here. 
An qualitative  estimate of the form of the $I-$band magnitude-mass relation for 
the Pleiades can be derived from the observations. For example, we
 start with the magnitude-mass relation from the 125Myr NEXTGEN model of Baraffe et
al. (1998) and assume this holds true as far as the top of the gap in
the sequence. We could equally use the models of Burrows et
al. (1997) or D'Antona \& Mazzitelli (1997), but for reasons outlined
in our previous work (e.g Jameson et al. 2002) we prefer the
calculation of the Lyon group. Equation 2 can be readily derived by 
integrating the MF from a mass m$_{1}$ to a lower mass m$_{2}$ and
again from m$_{2}$ to an even lower mass m$_{3}$, where $\alpha$ is the
index of a powerlaw model MF and N$_{12}$,N$_{23}$ the number of brown
dwarfs in the mass intervals m$_{1}$ to m$_{2}$ and m$_{2}$ to m$_{3}$ respectively. 

\vspace{0.1cm}
\begin{center}
\begin{equation}
N_{23}/N_{12} = (m_{2}^{1-\alpha} -m_{3}^{1-\alpha}) / (m_{1}^{1-\alpha} - m_{2}^{1-\alpha})
\end{equation}
\end{center} 
\vspace{0.1cm}

The ratio N$_{23}$/N$_{12}$ is determined from observation and m$_{1}$
and m$_{2}$ are taken from the NEXTGEN model just above the dip. 
Assuming a MF for the
Pleiades, dN/dm $\propto$ m$^{-\alpha}$, $\alpha\approx0.4$, as recently
derived by Jameson et al. (2002) over a broad mass range
(0.3M$_{\odot}\geq$m$\geq 0.035$M$_{\odot}$), Equation 2 can be solved
for m$_{3}$. Thus masses can be estimated down through the dip region
and below as a function of $I$, and hence the magnitude-mass relation derived. We find our estimates
to be relatively insensitive to the assumed index $\alpha$ unless
m$_{3}\ll $m$_{2}$.  The result for the Pleiades, using only those
candidate members with photometry consistent with them being single
objects, is shown in Figure~3 (open circles). Since the component
masses of the unresolved binary candidates are not known, a robust
treatment of these is difficult. However, if we assume all of these to
be equal mass binaries and the binary fraction above the gap, where
the data are incomplete, to be the same as below, then the
magnitude-mass relation is found to be nearly identical to that
obtained by considering only the single brown dwarfs, as above. We
choose not to proceed beyond $I=19.5$ as the survey is
complete over only a small area of sky at fainter magnitudes and
the statistics are poor. The overall result, as illustrated in Figure~3 by the dotted line, is that beyond the beginning of the gap any brown dwarf mass may be greater than predicted by either the NEXTGEN or the DUSTY models.

This procedure has been repeated for the $\sigma-$Orionis cluster,
using the $I-$band LF data reported in Bejar et al. (2001). Following our
argument above and since the single and binary sequences are not well
resolved in the Bejar et al. work, we have ignored the influence of
unresolved binarity, treating all their candidate members as
single. As the models of Burrows et al. (1997) are widely used to
predict the magnitudes of planetary mass objects we opt here to use, in
turn, both the 5Myr NEXTGEN model and the 5Myr evolutionary track of
the Arizona group as the starting points for this calculation. The latter
model has been transformed onto the observational plane using the
 bolometric corrections and T$_{\rm eff}$-colour relation of the former.
As shown in Figure~4 we have derived the magnitude-mass relation down
to $I=24.0$, although due to the completeness limit of the survey, the
point below $I=22$ provides  only an upper limit on mass. Except at
the lowest masses where the models and our semi-empirical estimates
appear to be reasonably consistent, we find once again that beyond the
beginning of the gap the mass of any brown dwarf may be greater than
predicted by the theoretical calculations.   
 
\section{Discussion}

Careful scrutiny of the evidence cited above suggests that the width
of the LF dip is greater in the younger $\sigma-$ and $\theta-$Orionis
clusters than in either the Pleiades or the field populations. This is
consistent with our interpretation in which the luminosity-mass
relation undergoes a sharp change in form between spectral types
M7-M8. For a population older than 100Myrs, radius is virtually
independent of mass right across the planetary/brown dwarf mass
regime. However, in a population with an age of only a few Myrs,
radius decreases rapidly with mass. Therefore, as one moves to lower
masses and T$_{\rm eff}$s, luminosity decreases more rapidly in
younger populations, and one should find a wider dip in the LF,
representative of the dip spanning a constant range in T$_{\rm eff}$. 
From Figure~11 in Burrows et al. (1997) we estimate that the dip in 
the LF should be $\sim 2-2.5\times$ wider (in magnitudes) in a 5 Myr
population than in a 125Myr population, in agreement with the
observations.  
 
We now speculate on the possible cause of a sharp local change in the
form of the luminosity-mass relation in this T$_{\rm eff}$ regime. We
note that a dip in the LF observed in the sequences of both open
clusters and the field at M$_{V}\approx7$ (the Wielen gap; e.g Wielen,
Jahreiss \& Kruger 1983; Belikov at al. 1998), has been attributed to
a local steepening of the luminosity-mass relation due to strong
H$^{-}$ opacity in the atmospheres of objects in this T$_{\rm eff}$
regime (Kroupa, Tout \& Gilmore 1990). Theoretical cool model
atmospheres predict that dust grains of Al$_{2}$O$_{3}$, Fe and
MgSiO$_{4}$ begin to condense in the outermost layers of a brown dwarf
(or very low mass star) at T$\sim1800$K, corresponding to a T$_{\rm
eff}$ consistent with the M7-M8 LF dip (e.g Allard et al. 2001, Tsuji
2001). Indeed, spectroscopic observations of late-M field dwarfs lend
support to these predictions (Jones \& Tsuji 1997). However, the DUSTY
evolutionary models of Chabrier et al. (2000) which include a
treatment for the formation of dust, do not predict any discontinuity
in the luminosity-mass relation here. In the synthetic atmospheres of
Allard et al., which provide the outer boundary conditions for the
DUSTY calculations, dust is treated as grains with sizes in the range
$0.00625-0.24\mu$m. In this case, dust does not contribute
significantly to the opacities in the near-infrared where the water
bands are dominant (Allard et al. 2001). However, as illustrated by
Figure~4 of Allard et al. if the grain size is increased into the
regime described by Mie scattering theory, the opacity provided by dust is dramatically enhanced (we note
that the opacity of dust increases with decreasing wavelength). While current observations argue against grains as large as $100\mu$m, the most detailed current cloud formation models predict sizes in the range $5-20\mu$m (Cooper et al. 2002). It might well be possible to reproduce the observed dips in the LFs with a moderate increase in the model grain size.

With the recent advent of wide-field near-infrared imagers (e.g
FLAMINGOS) and the ever increasing availability of wide-field optical
imagers (e.g WFI) the results of deeper, larger area surveys for low
mass members of rich young clusters such as NGC2547, NGC2516 and
IC2391 should soon become available. The improved statistics of these
surveys will confirm or otherwise the reported dip in the LF between M7-M8 and allow a more thorough examination of its implications for mass determinations at T$_{\rm eff}\simless2700$K. 

\section{Summary}

1. We have presented evidence for a deficit of very-low mass stars/brown dwarfs between spectral types M7-M8 and have argued that this arises from a sharp local change in the shape of the luminosity-mass relation.

\vspace{0.1cm}

\noindent 2. Structure near the bottom end of the
Trapezium IMF recently proposed by Muench et al. (2002) is more likely
a manifestation of
the feature in the luminosity-mass relation reported here.

\vspace{0.1cm}

\noindent 3. We find that brown dwarfs with spectral types later than M7-M8 may have masses greater than predicted by current theoretical models. 

\vspace{0.1cm}

\noindent 4. We speculate that the onset of dust formation in the upper atmosphere of objects in this T$_{\rm eff}$ regime is responsible for the change in the shape of the luminosity-mass relation. This might be modelled by assuming larger grain sizes than are currently used in the DUSTY calculations.

\section*{Acknowledgments}
PDD, DJP and STH acknowledge the financial support of PPARC. We thank
Gilles Chabrier and Isabelle Baraffe for useful discussions on their
evolutionary models, David Barrado y Navascués for making his latest
results available to us and the referee, Nigel Hambly, for a swift and
useful response.

\bsp

\label{lastpage}

\end{document}